\documentclass[fleqn,usenatbib,useAMS]{mnras}
 \usepackage{graphicx}
 \usepackage{amsmath}
 \usepackage{amssymb}
 \usepackage[T1]{fontenc}
 \usepackage{ae,aecompl}
 \usepackage{txfonts}
 \usepackage{enumerate}

 \title[Dynamically correlated comets and TNOs]
       {Dynamically correlated minor bodies in the outer Solar system}

 \author[C. de la Fuente Marcos and R. de la Fuente Marcos]
        {C.~de~la~Fuente~Marcos\thanks{E-mail: nbplanet@ucm.es}
         and
         R. de la Fuente Marcos \\
         Universidad Complutense de Madrid,
         Ciudad Universitaria, E-28040 Madrid, Spain}
 \date{Accepted 2017 October 20.
       Received 2017 October 20;
       in original form 2017 September 4}
 \pubyear{2018}
 
\begin{document}
  \label{firstpage}
  \pagerange{\pageref{firstpage}--\pageref{lastpage}}
  \maketitle

  \begin{abstract}
     The organization of the orbits of most minor bodies in the Solar system 
     seems to follow random patterns, the result of billions of years of 
     chaotic dynamical evolution. Much as heterogeneous orbital behaviour is 
     ubiquitous, dynamically coherent pairs and groups of objects are also
     present everywhere. Although first studied among the populations of 
     asteroids and comets that inhabit or traverse the inner Solar system, 
     where they are very numerous, at least one asteroid family has been 
     confirmed to exist in the outer Solar system and two other candidates 
     have been proposed in the literature. Here, we perform a systematic 
     search for statistically significant pairs and groups of dynamically 
     correlated objects through those with semimajor axis greater than 25~au, 
     applying a novel technique that uses the angular separations of orbital 
     poles and perihelia together with the differences in time of perihelion 
     passage to single out pairs of relevant objects. Our analysis recovers 
     well-known, dynamically coherent pairs and groups of comets and 
     trans-Neptunian objects and uncovers a number of new ones, prime 
     candidates for further spectroscopic study. 
  \end{abstract}


  \section{Introduction}
     Collisional break-ups, rotational or thermal-stress-induced splittings, tidal disruptions and binary dissociations all lead to the 
     formation of pairs or groups of dynamically correlated minor bodies (see e.g. Benz \& Asphaug 1999; Boehnhardt 2004; Sekanina \& Chodas 
     2005, 2007; Bottke et al. 2006; Jacobson \& Scheeres 2011; Schunov{\'a} et al. 2014; Jacobson 2016; Vokrouhlick{\'y} et al. 2017a); 
     mean motion and secular resonances can also induce orbital coherence (see e.g. de la Fuente Marcos \& de la Fuente Marcos 2016a). 
     Groups or families were first identified in the main asteroid belt (Hirayama 1918), but they are present in other regions of the Solar 
     system as well (Brown et al. 2007). As many as 100\,000 asteroids have been found to be members of families (see e.g. Nesvorn\'y, 
     Bro\v{z} \& Carruba 2015), but dozens of comets are also organized in families (see e.g. Sekanina \& Chodas 2005). 
     
     The largest asteroid families ---Nysa, Vesta, Flora and Eos--- may include tens of thousands of members (see e.g. Nesvorn\'y et al. 
     2015), the smallest ---e.g. Datura (Nesvorn\'y, Vokrouhlick{\'y} \& Bottke 2006; Nesvorn\'y \& Vokrouhlick{\'y} 2006; Vokrouhlick{\'y} 
     et al. 2009; Vokrouhlick{\'y} et al. 2017b; Rosaev \& Pl{\'a}valov{\'a} 2017; Henych \& Holsapple 2018), Lucascavin and Emilkowalski 
     (Nesvorn\'y \& Vokrouhlick{\'y} 2006), or Haumea (Brown et al. 2007; Ragozzine \& Brown 2007; Snodgrass et al. 2010; Carry et al. 
     2012)--- may host a few tens or less. Unbound pairs of asteroids, probably of a common origin, have also been identified 
     (Vokrouhlick{\'y} \& Nesvorn{\'y} 2008; Pravec et al. 2010; Jacobson 2016); one candidate pair resides in the scattered disc 
     (Rabinowitz et al. 2011). Groups of pairs define young asteroid clusters (Pravec et al. 2018). Although there are hundreds of asteroids 
     known to have one or more moons (see e.g. Margot et al. 2015), binary comets seem to be uncommon; there is only one confirmed example, 
     (300163) 2006~VW$_{139}$=288P (Agarwal et al. 2017), but comets with bilobate nuclei like 8P/Tuttle (Harmon et al. 2010) or 
     67P/Churyumov-Gerasimenko (Massironi et al. 2015) are known to exist. Nonetheless, several pairs of comets having nearly identical 
     orbits have been detected (see e.g. Sekanina \& Kracht 2016). 

     The first bona fide asteroid family identified in the outer Solar system was the one associated with dwarf planet Haumea (Brown et al. 
     2007). Predating this discovery by a few years, a candidate collisional family was proposed by Chiang (2002). The subject of finding 
     collisional families of trans-Neptunian objects (TNOs) has been studied by Chiang et al. (2003) and Marcus et al. (2011). Here, we 
     perform a systematic search for statistically significant pairs and groups of dynamically correlated objects through those with 
     semimajor axis greater than 25~au, applying a novel technique that uses the angular separations of orbital poles and perihelia together 
     with the differences in time of perihelion passage to single out pairs of relevant objects from which groupings can eventually be 
     uncovered. This paper is organized as follows. Section~2 reviews the cases of the comets 73P/Schwassmann-Wachmann~3 and D/1993 F2 
     (Shoemaker-Levy 9), and the Haumea collisional family. The distributions of relevant parameters at various distances from the Sun are 
     explored in Section~3. Some statistically significant pairs and groups found by our approach are presented and discussed in Section~4. 
     In Section~5, we summarize our conclusions. 

  \section{What to expect after disruption}
     The first step in searching for dynamically correlated minor bodies, particularly those resulting from break-ups, is to get a clear 
     characterization of what the expectations may be. The outcome of cometary disruption is well documented through two well-studied 
     examples, those of the comets 73P/Schwassmann-Wachmann~3 and D/1993~F2 (Shoemaker-Levy~9). For reasons that still remain unclear, comet 
     73P started to break apart in 1995 and dozens of fragments were observed in 2006 and 2007 (see e.g. Crovisier et al. 1996; Weaver et 
     al. 2006; Reach et al. 2009; Hadamcik \& Levasseur-Regourd 2016). Some of these fragments have been recovered in 2010--2011 (Harker et 
     al. 2011, 2017; Sitko et al. 2011) and 2016--2017 (e.g. Kadota et al. 2017;\footnote{\url{http://www.minorplanetcenter.net/mpec/K17/K17C79.html}} 
     Williams 2017); it may consist of hundreds of pieces now (68 of them have orbit determinations). This fragmentation process can be 
     described as gentle and progressive. In striking contrast, comet Shoemaker-Levy~9 experienced a sudden, violent fragmentation event 
     triggered by strong tidal forces during a close encounter with Jupiter in 1992 July (see e.g. Sekanina, Chodas \& Yeomans 1994, 1998; 
     Asphaug \& Benz 1996; Sekanina 1997). Most fragments collided with Jupiter over a period of a week (1994 July 16--22); 21 of them have 
     orbit determinations. Quite different may have been the collisional event that led to the formation of the Haumea family (Brown et al. 
     2007; Schlichting \& Sari 2009; Leinhardt, Marcus \& Stewart 2010; Lykawka et al. 2012; Ortiz et al. 2012) perhaps more than 1~Gyr ago 
     (Ragozzine \& Brown 2007; Volk \& Malhotra 2012). Fragments of recently disrupted minor bodies must have very similar values of their 
     semimajor axis, $a$, eccentricity, $e$, inclination, $i$, longitude of the ascending node, $\mathit{\Omega}$, argument of perihelion, 
     $\omega$, and time of perihelion passage, $\tau_{q}$, but $\mathit{\Omega}$, $\omega$ and $\tau_{q}$ tend to become increasingly 
     randomized over time. In contrast, recently unbound pairs resulting from binary dissociation events might have relatively different 
     values of $a$ and $e$, but very similar values of $i$, $\mathit{\Omega}$ and $\omega$, the difference in $\tau_{q}$ may initially range 
     from weeks to centuries, but grows rapidly over time (see e.g. de Le{\'o}n, de la Fuente Marcos \& de la Fuente Marcos 2017; de la 
     Fuente Marcos, de la Fuente Marcos \& Aarseth 2017).  

     In this paper, we use the dynamical signatures left by the three fragmentation events mentioned before ---in the form of distributions 
     of possible values of certain parameters--- to single out dynamically coherent pairs and groups of minor bodies. We focus on the values 
     of the angular separations between the poles, $\alpha_{\rm p}$, and perihelia, $\alpha_q$, of the orbits of the members of the pair 
     calculated as described by e.g. de la Fuente Marcos \& de la Fuente Marcos (2016b), and the differences in their times of perihelion 
     passage, $\Delta\tau_{q}$. We start by computing the distributions of $\alpha_{\rm p}$, $\alpha_q$ and $\Delta\tau_{q}$ for the objects 
     associated with those three fragmentation episodes, but the procedure described here is applied in other sections as well. Our 
     calculations use the values of $i$, $\mathit{\Omega}$, $\omega$, and $\tau_{q}$ of real objects, and their respective standard 
     deviations, $\sigma_i$, $\sigma_{\mathit{\Omega}}$, $\sigma_{\omega}$ and $\sigma_{\tau_{q}}$. The source of the data in our table and
     figures is Jet Propulsion Laboratory's Solar System Dynamics Group Small-Body Database (JPL's SSDG SBDB, Giorgini, Chodas \& Yeomans 
     2001; Giorgini 2011, 2015),\footnote{\url{https://ssd.jpl.nasa.gov/sbdb.cgi}} and we restrict the analysis to objects with 
     $\sigma_a/a$$<$0.05 (data as of 2017 October 3), where $\sigma_a$ is the value of the standard deviation of $a$. In order to produce 
     the distributions of $\alpha_{\rm p}$, $\alpha_q$ and $\Delta\tau_{q}$, we generate 10$^{7}$ random pairs of virtual objects and 
     compute for each one of them the values of $\alpha_{\rm p}$, $\alpha_q$ and $\Delta\tau_{q}$. The orbit of each random virtual object 
     is calculated using the means and standard deviations of the orbit determinations of real objects. For example, in order to compute a 
     new random value of $\mathit{\Omega}$, the expression $\mathit{\Omega}_{\rm r} = \langle{\mathit{\Omega}}\rangle + 
     \sigma_{\mathit{\Omega}}\,r_{\rm i}$ is used, where $\mathit{\Omega}_{\rm r}$ is the longitude of the ascending node of the random 
     orbit, $\langle{\mathit{\Omega}}\rangle$ is the mean value of the longitude of the ascending node of one real object, 
     $\sigma_{\mathit{\Omega}}$ is its associated standard deviation, and $r_{\rm i}$ is a (pseudo-) random number with a normal 
     distribution in the range $-$1 to 1. The same procedure is utilized for the other orbital elements. Each virtual pair has been 
     generated using the means and standard deviations of the orbit determinations of two different real objects.

\begin{figure*}
        \centering
         \includegraphics[width=0.33\linewidth]{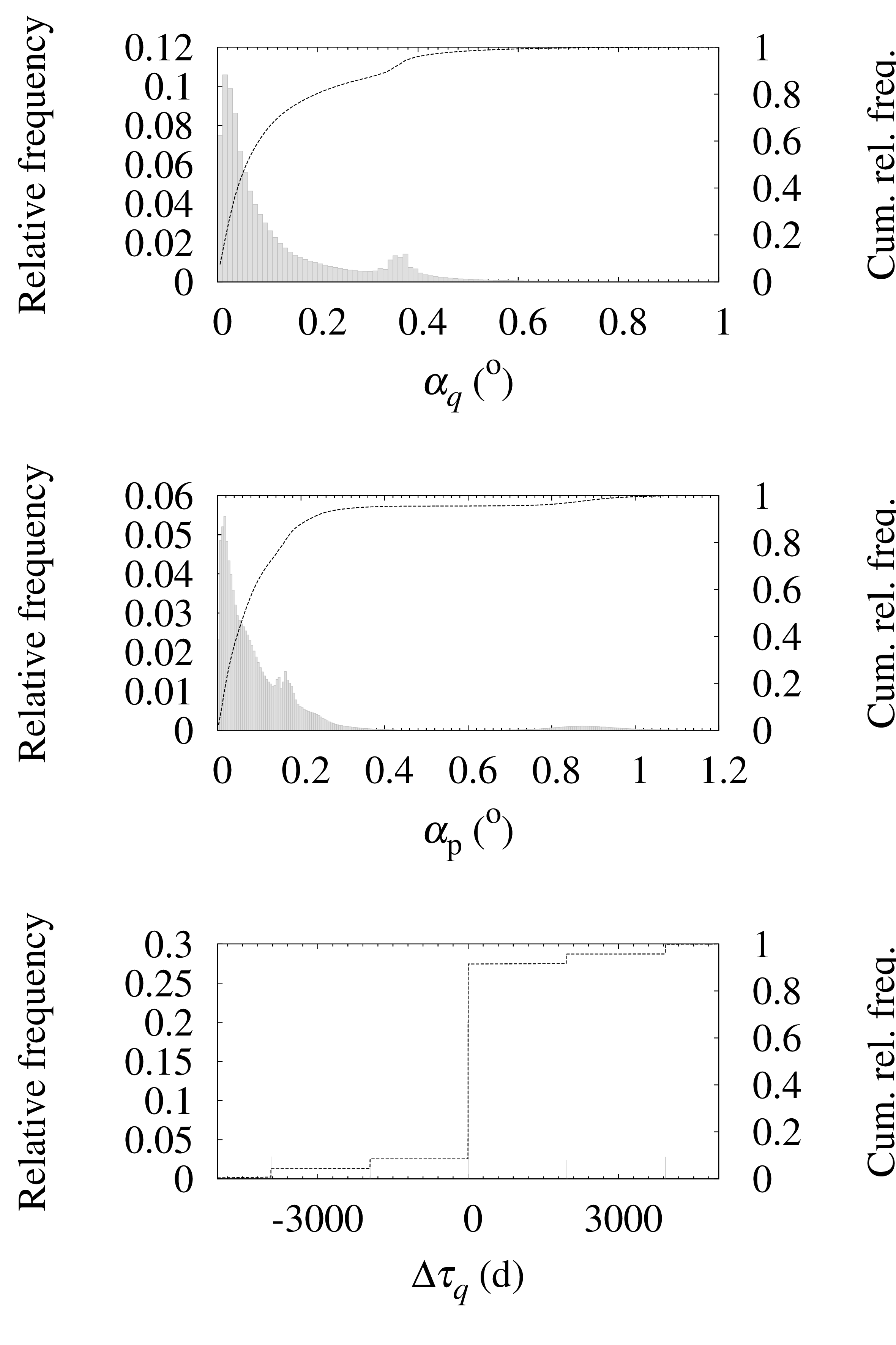}
         \includegraphics[width=0.33\linewidth]{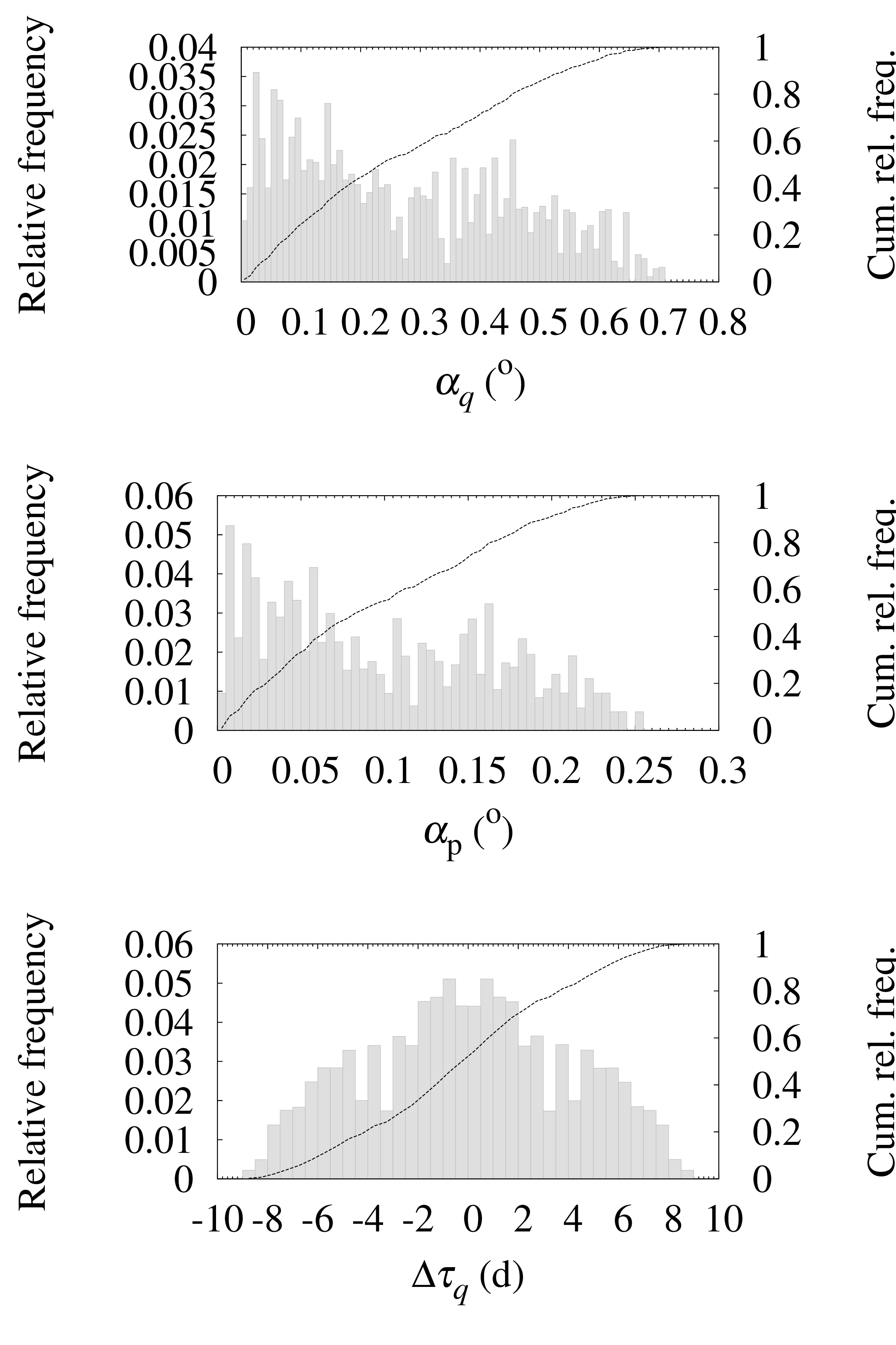}
         \includegraphics[width=0.33\linewidth]{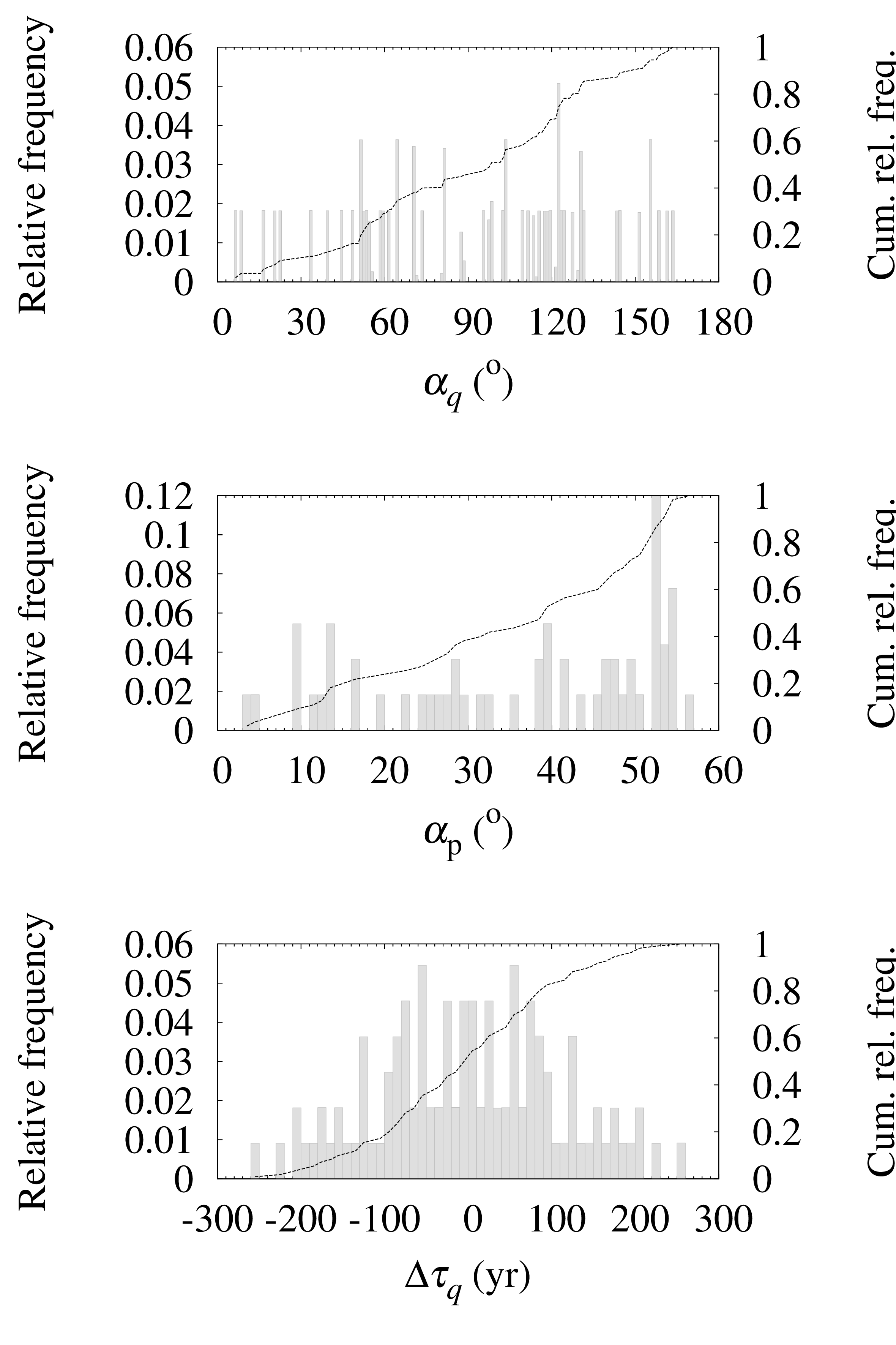}
         \caption{Distributions of possible angular separations between perihelia (top panels) and poles (middle panels), and differences in
                  time of perihelion passage (bottom panels) of the orbits of pairs of fragments of the comets 73P/Schwassmann-Wachmann~3 
                  (left-hand panels) and D/1993 F2 (Shoemaker-Levy 9) (central panels), and members of the Haumea collisional family 
                  (right-hand panels).  
                 }
         \label{disruption}
      \end{figure*}

     Fig.~\ref{disruption} shows the distributions of possible angular separations, $\alpha_{\rm p}$ and $\alpha_q$, and also of 
     $\Delta\tau_{q}$ for the outcomes of the three events mentioned before. The dynamical signature associated with comet 73P 
     (Fig.~\ref{disruption}, left-hand panels) has been computed using 45 fragments, those with $\sigma_a/a$$<$0.05. A disruption episode 
     that is taking place over an extended period of time, two decades, leaves a signature with $\alpha_{\rm p}$$<$1{\degr} and 
     $\alpha_q$$<$0\fdg6; the distribution in $\Delta\tau_{q}$ reflects the fact that different groups of fragments have been observed at 
     different epochs and that these fragments have been released at various times. The distributions obtained for the 21 fragments of 
     Shoemaker-Levy~9 (Fig.~\ref{disruption}, central panels) are mostly consistent with those of comet 73P, $\alpha_{\rm p}$$<$0\fdg25 and 
     $\alpha_q$$<$0\fdg7, with $|\Delta\tau_{q}|$$<$10~d. These results show that any fragments produced by a relatively recent disruption 
     event must have low values of $\alpha_{\rm p}$ and $\alpha_q$, probably under $\sim$2\degr; on the other hand, the value of 
     $|\Delta\tau_{q}|$ must be significantly shorter than the average orbital period of the fragments. In striking contrast, fragments from 
     an old disruption episode may have values of $\alpha_q$ uniformly distributed in the interval (0,~180)\degr, values of 
     $|\Delta\tau_{q}|$ spanning the entire relevant orbital period of the fragments, and a wide range in the values of $\alpha_{\rm p}$ as 
     observed in the case of the Haumea family (Fig.~\ref{disruption}, right-hand panels). The distributions for the Haumea family have 
     been computed using the following TNOs (see e.g. Thirouin et al. 2016): (136108)~Haumea 2003~EL$_{61}$, (24835) 1995~SM$_{55}$, (19308) 
     1996~TO$_{66}$, (86047) 1999~OY$_{3}$, (55636) 2002~TX$_{300}$, (120178) 2003~OP$_{32}$, 2003~SQ$_{317}$, (416400) 2003~UZ$_{117}$, 
     (308193) 2005~CB$_{79}$, (145453) 2005~RR$_{43}$ and (386723) 2009~YE$_{7}$. The average values and standard deviations of $a$, $e$,
     $i$, $\mathit{\Omega}$ and $\omega$ for the Haumea collisional family (11 assumed members) are 43.2$\pm$0.7~au, 0.13$\pm$0.03, 
     27\fdg5$\pm$1\fdg4, 184\degr$\pm$105\degr and 197\degr$\pm$98\degr, i.e. the values of $\mathit{\Omega}$ and $\omega$ are consistent 
     with those from a uniform distribution, 180\degr$\pm$104\degr. Dwarf planet Haumea also hosts two moons ---Hi`iaka (Brown et al. 2005) 
     and Namaka (Brown et al. 2006)--- and one ring (Ortiz et al. 2017). Hereinafter, our approach assumes that the values of the relevant 
     angular separations of any pair of interest will resemble those in Fig.~\ref{disruption}, left-hand and central panels; this 
     assumption can only lead to uncover very recent (in astronomical terms) disruption events. The pair of values $\alpha_{\rm p}$ and 
     $\alpha_q$ is used as a proxy to assess the degree of dynamical coherence, i.e. the lower the values, the higher the level of 
     coherence. The value of $\Delta\tau_{q}$ is utilized to estimate the dynamical age.
%
%
%
%
  \section{Relevant global distributions}
\begin{figure*}
        \centering
         \includegraphics[width=0.33\linewidth]{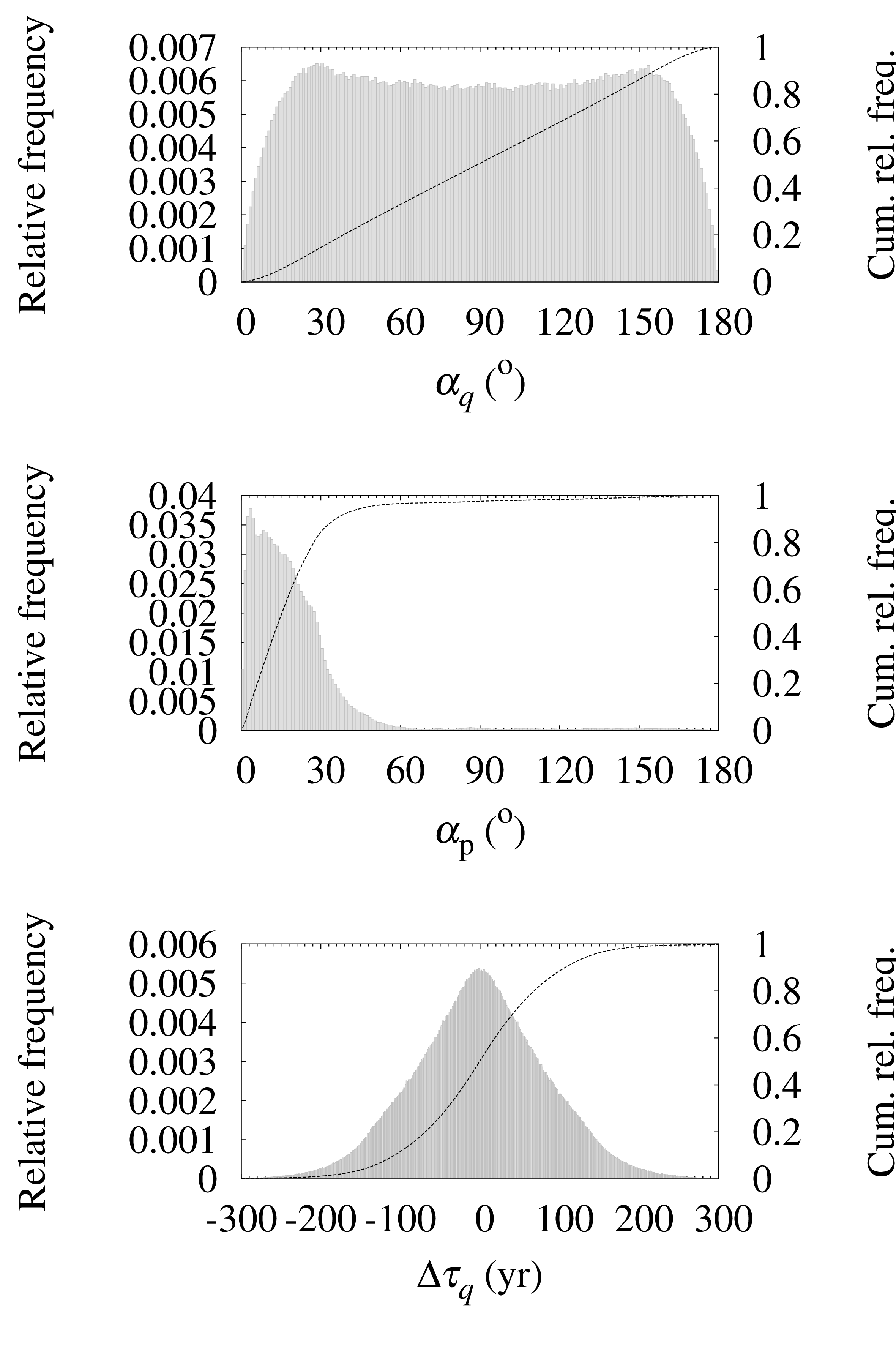}
         \includegraphics[width=0.33\linewidth]{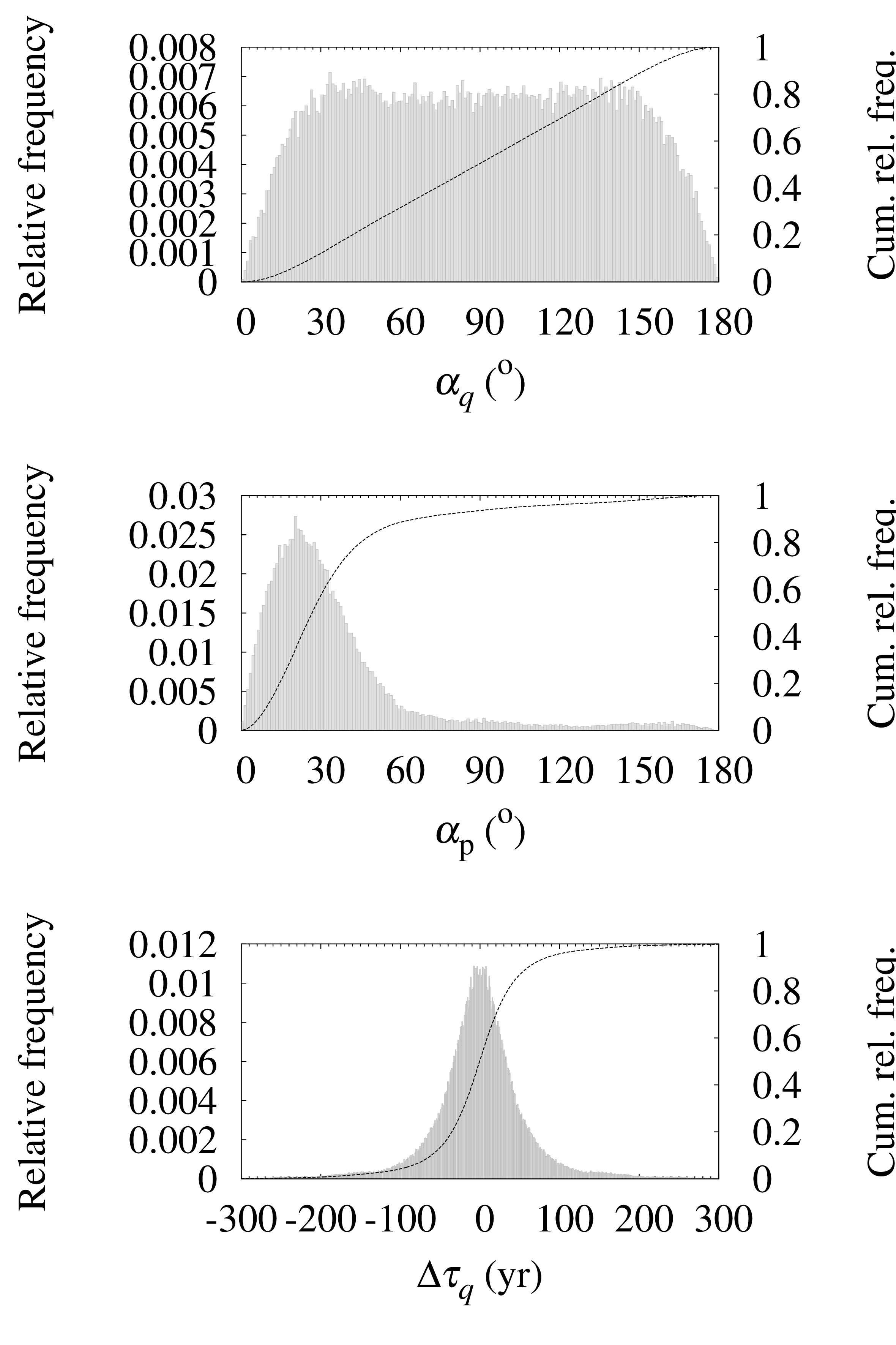}
         \includegraphics[width=0.33\linewidth]{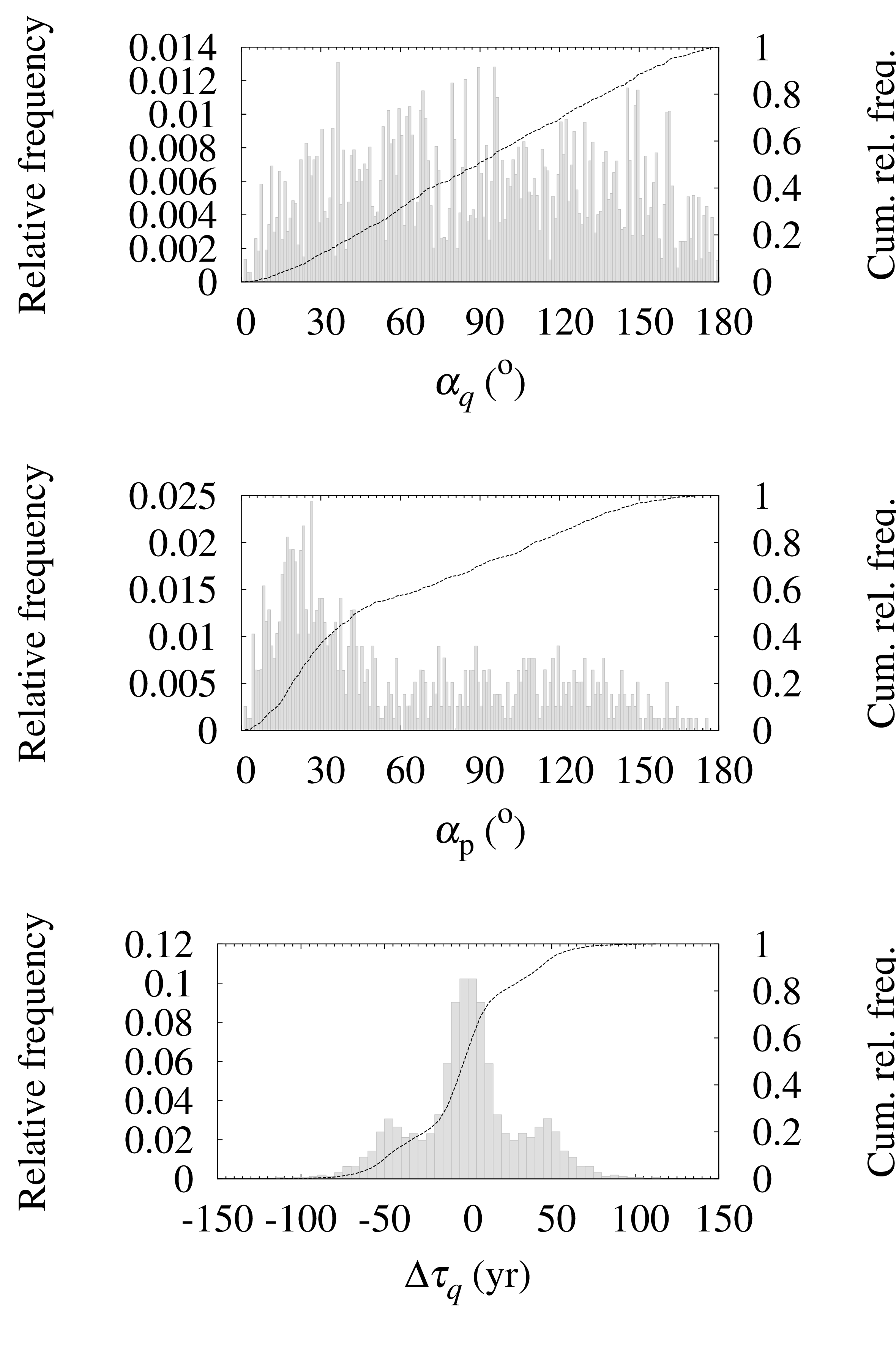}
         \caption{Distributions of possible angular separations between perihelia (top panels) and poles (middle panels), and differences in
                  time of perihelion passage (bottom panels) of the orbits of pairs of minor bodies with semimajor axis in the ranges 
                  25--50~au (left-hand panels), 50--150~au (central panels), and extreme Centaurs and trans-Neptunian objects 
                  ($a$$>$150~au, right-hand panels). The bin size is 1{\degr} in both top and middle panels; the bottom panels have bin 
                  sizes of 1 yr (left-hand and central panels) and 5 yr (right-hand panel).
                 }
         \label{distributions}
      \end{figure*}

\begin{figure}
        \centering
         \includegraphics[width=\linewidth]{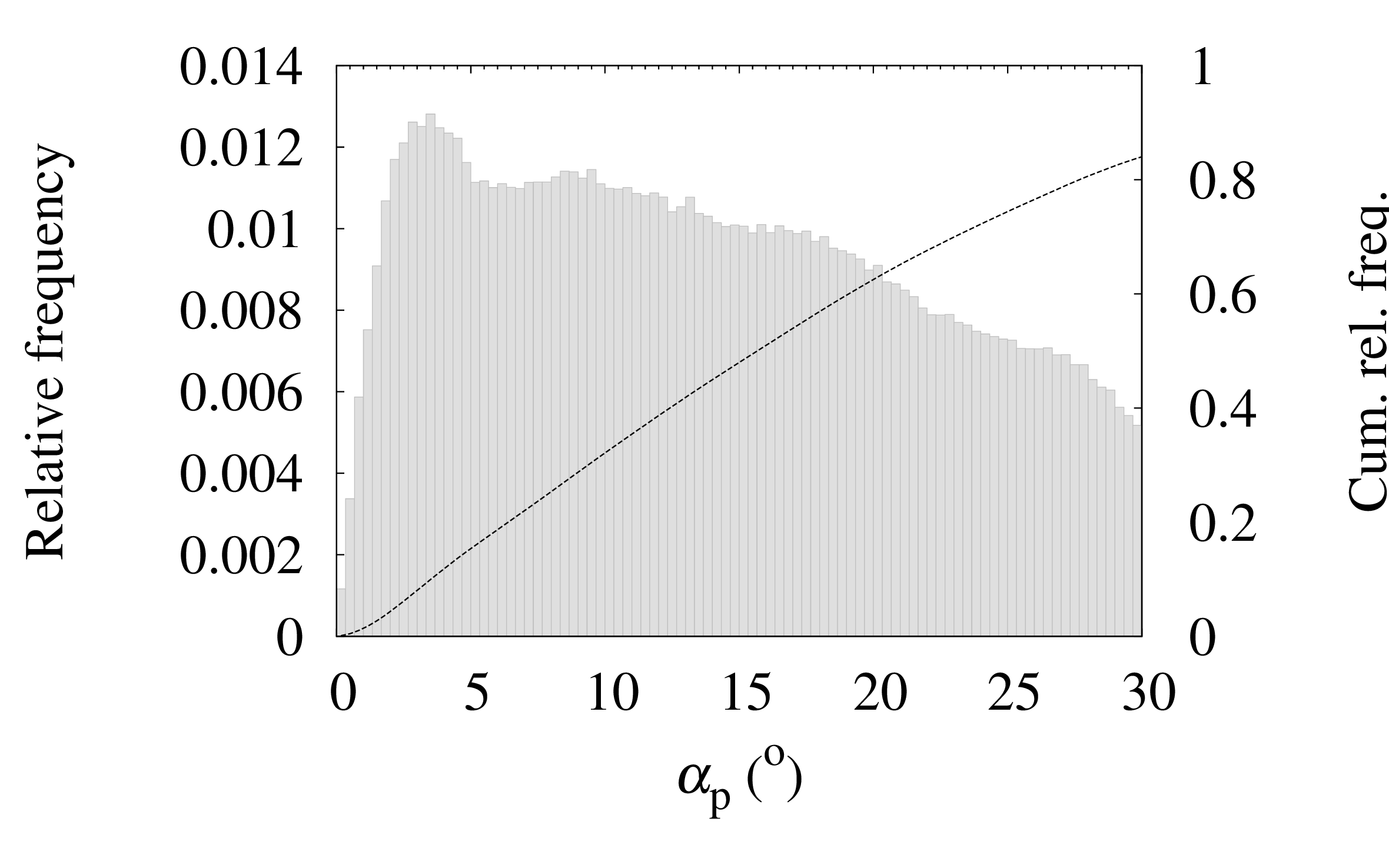}
         \caption{Detail of the angular separations between poles for pairs of minor bodies with semimajor axis in the range 25--50~au 
                  (magnified version of Fig.~\ref{distributions}, left-hand, middle panel). The bin size is 1/3\degr.
                 }
         \label{distributionsLmz}
      \end{figure}

     In order to single out statistically significant pairs of dynamically correlated objects, we have computed the distributions of 
     $\alpha_{\rm p}$, $\alpha_q$ and $\Delta\tau_{q}$ at various distance ranges using the procedure described in the previous section. 
     Fig.~\ref{distributions} is conceptually similar to Fig.~\ref{disruption}, but now larger sets of mostly unrelated objects are used; 
     minor bodies with semimajor axis in the range 25--50~au (1358 objects) are analysed in the left-hand panels, those with values in 
     the range 50--150~au (335) are plotted in the central panels, and those with $a$$>$150~au (40) appear in the right-hand panels. 
     For the range 25--50~au, the probability of finding a pair with both $\alpha_{\rm p}$ and $\alpha_q$ under 2{\degr} is $3.8 \times 
     10^{-4}$, at 50--150~au is $4.5 \times 10^{-5}$, and beyond 150~au is $<$$10^{-7}$ (probabilities calculated in the usual way; see e.g. 
     Wall \& Jenkins 2012). These very low $p$-values mean that pairs of dynamically coherent minor bodies in the outer Solar system are 
     rare, but also that the ones found are probably statistically significant ---i.e. not due to chance--- with few interlopers. The 
     distribution of possible angular separations between poles (Fig.~\ref{distributions}, middle panels) shows two peaks for TNOs with 
     semimajor axis in the range 25--50~au (left-hand, middle panel). The primary maximum at about 3{\degr} corresponds to the cold 
     population described by e.g. Petit et al. (2011), while the secondary one at about 9{\degr} (far less obvious than the primary one) 
     signals the transition to the hot population (see Fig.~\ref{distributionsLmz}). The peaks at $a\in(25, 50)$~au are well below the 
     single peak observed for TNOs with $a\in(50, 150)$~au at nearly 21{\degr} (Fig.~\ref{distributions}, central, middle panel) and its 
     relative frequency is lower; data for $a$$>$150~au are still very incomplete, but the peak appears to be closer to 30{\degr} 
     (Fig.~\ref{distributions}, right-hand, middle panel) and its relative frequency lower than those of the inner TNOs (see 
     Fig.~\ref{distributions}, left-hand and central, middle panels). The three cumulative distributions are very different, with that 
     of TNOs with $a$$>$150~au showing the effects of a significant fraction of retrograde orbits (17.5 per cent), which are relatively 
     scarce for TNOs with $a$$<$150~au (1.7 per cent). The origin of the large difference in the fraction of retrograde orbits remains 
     unclear, but the presence of one or more yet-to-be-discovered planetary bodies orbiting the Sun well beyond Neptune may be able to 
     explain such dissimilarity (see e.g. de la Fuente Marcos, de la Fuente Marcos \& Aarseth 2015, 2016; Batygin \& Brown 2016).
%
%
%
%
%
%
%
%
     
\begin{figure}
        \centering
         \includegraphics[width=\linewidth]{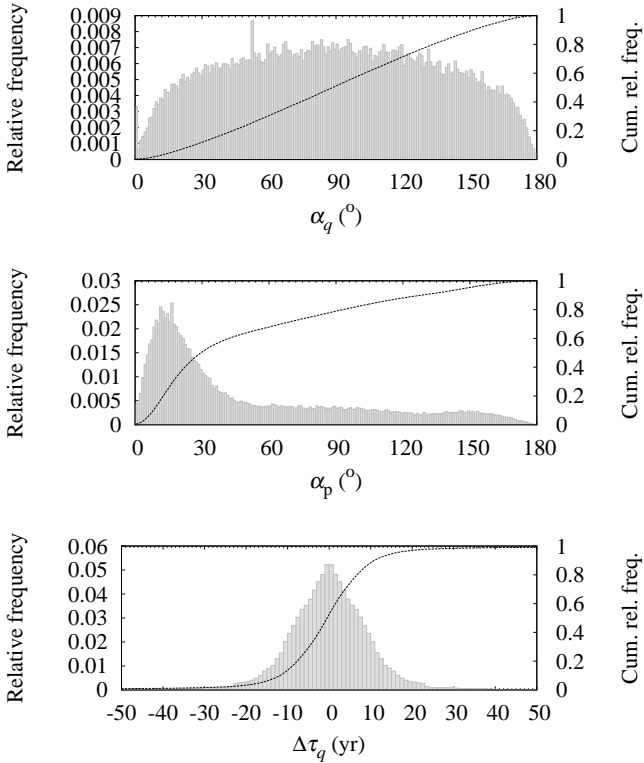}
         \caption{Same as Fig.~\ref{distributions} but for comets with $a$$<$1000~au. The spikes in the histograms of $\alpha_{\rm p}$ and 
                  $\alpha_q$ are due to the presence of multiple sets of comet fragments, for example those of comets 
                  73P/Schwassmann-Wachmann~3, D/1993~F2 (Shoemaker-Levy~9) and 332P/Ikeya-Murakami. The bin size is 1{\degr} in both top and 
                  middle panels and 1 yr for the bottom panel.
                 }
         \label{distributionsC}
      \end{figure}

     Our approach was able to recover the following comets (distributions shown in Fig.~\ref{distributionsC}), all known to have 
     fragments: 51P/Harrington (Kidger \& Manteca 2002), 57P/duToit-Neujmin-Delporte, 128P/Shoemaker-Holt~1, 141P/Machholz~2 (Sekanina 
     1999), 205P/Giacobini, 213P/Van Ness, 332P/Ikeya-Murakami (Jewitt et al. 2016; Kleyna et al. 2016), C/2003~S4 (LINEAR), P/2004~V5 
     (LINEAR-Hill), P/2013~R3 (Catalina-PANSTARRS) (Jewitt et al. 2017) and P/2016~J1 (PANSTARRS) (Hui, Jewitt \& Du 2017b). In addition, we 
     recovered the pair C/2002~A1 (LINEAR) and C/2002~A2 (LINEAR) discussed by Sekanina et al. (2003), the pair 169P/NEAT and P/2003~T12 
     (SOHO) found by Sosa \& Fern{\'a}ndez (2015), the pair C/1996~Q1 (Tabur) and C/2015~F3 (SWAN) identified by Sekanina \& Kracht (2016), 
     and a number of previously unknown ones (non-exhaustive list), 10P/Tempel~2 and P/2015~T3 (PANSTARRS), 208P/McMillan and P/2011~Q3 
     (McNaught), 342P/SOHO and P/2002~S7 (SOHO), 16P/Brooks~2 and 307P/LINEAR, and 285P/LINEAR and P/2013~N5 (PANSTARRS). Of particular 
     interest could be P/2010~B2 (WISE) that appears to be related to the multiple fragments of comet 332P/Ikeya-Murakami, confirming the 
     analysis carried out by Hui, Ye \& Wiegert (2017a). All these pairs and groups of fragments have very small values of both 
     $\alpha_{\rm p}$ and $\alpha_q$ ($<$2\degr) and in some cases very small values of $\Delta\tau_{q}$ (less than few months). The 
     probability of finding two comets with $a$$<$1000~au and both $\alpha_{\rm p}$ and $\alpha_q$ under 2{\degr} is about 0.0031; if we add 
     the constraint of having $|\Delta\tau_{q}|$$<$1~yr, the probability is 0.0026 that we interpret as strong evidence that the vast 
     majority of orbitally coherent comets must be the result of very recent splitting events. The data currently available suggest that it 
     is easier to find dynamically correlated candidate pairs among comets than among asteroids and that comets do not tend to keep 
     fragments gravitationally bound after disruption (i.e. binary or higher multiplicity comets are truly uncommon). On the other hand, 
     there is at least one example of a pair of very long-period comets, C/1988~F1 (Levy) and C/1988~J1 (Shoemaker-Holt), with relative 
     differences in $a$, $e$, $i$, $\mathit{\Omega}$, $\omega$ and $\tau_q$, as low as 8.7~au, 0.00006, 0\fdg0033, 0\fdg0015, 0\fdg0073 and 
     76.33~d, respectively (Sekanina \& Kracht 2016); such comets must be genetically related and the fragmentation event that led to them 
     must have happened at hundreds of astronomical units from the Sun. Regarding the distribution of possible angular separations between 
     poles, Fig.~\ref{distributionsC}, middle panel, shows a not-well-defined peak at about 15{\degr}. As a closing remark for this section, 
     our approach is confirmed to be statistically robust as multiple, well-established results are reproduced. 
%
%
%
%

  \section{Some statistically significant pairs}
     Here, we apply the procedure described in the previous sections to find pairs and groups of dynamically correlated objects in the outer 
     Solar system. Table~\ref{candidates} only shows our most significant findings, a more exhaustive analysis will be presented in a future 
     paper.

     \subsection{Semimajor axes in the range 25--50~au}
\begin{figure*}
        \centering
         \includegraphics[width=0.33\linewidth]{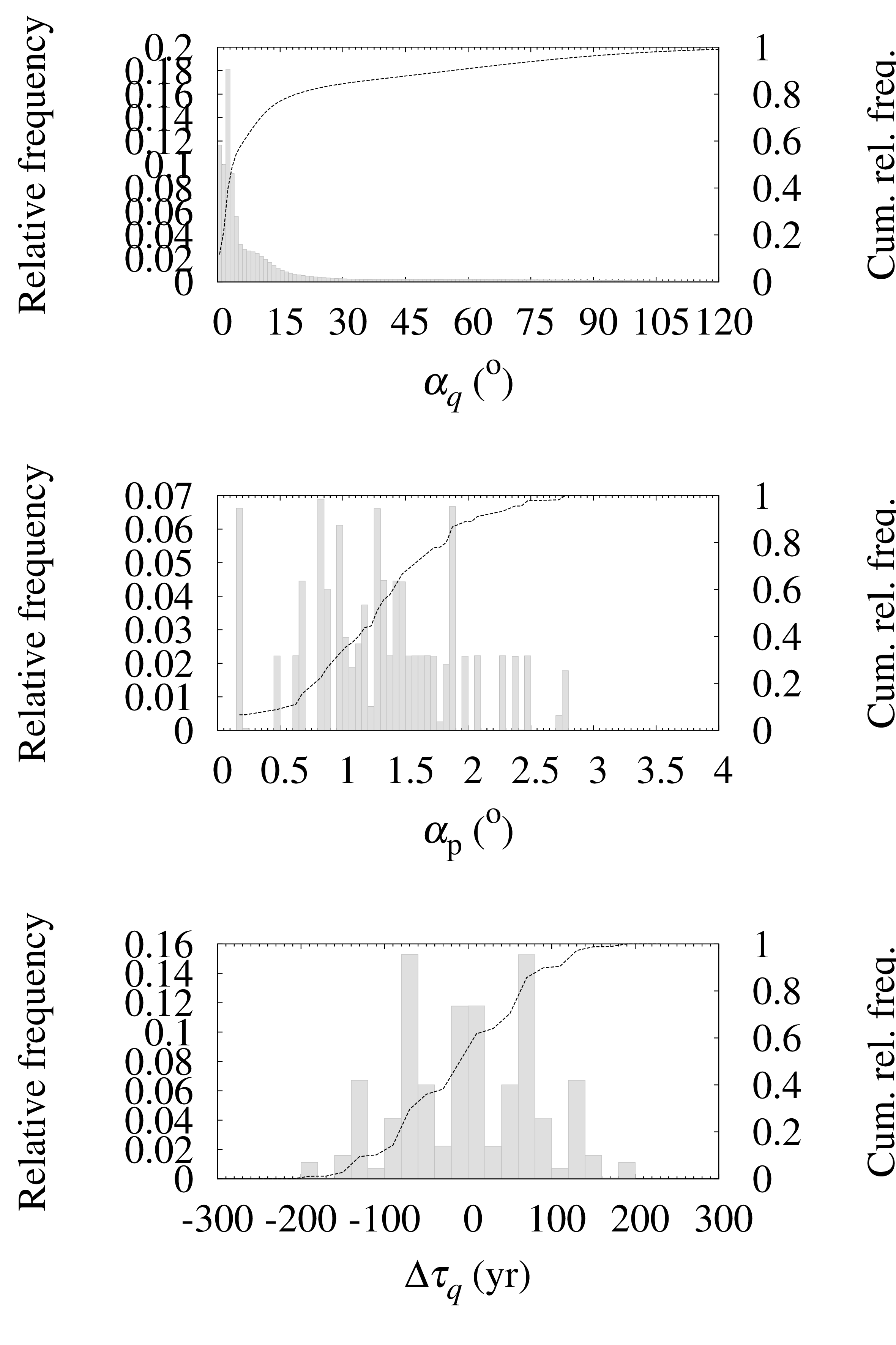}
         \includegraphics[width=0.33\linewidth]{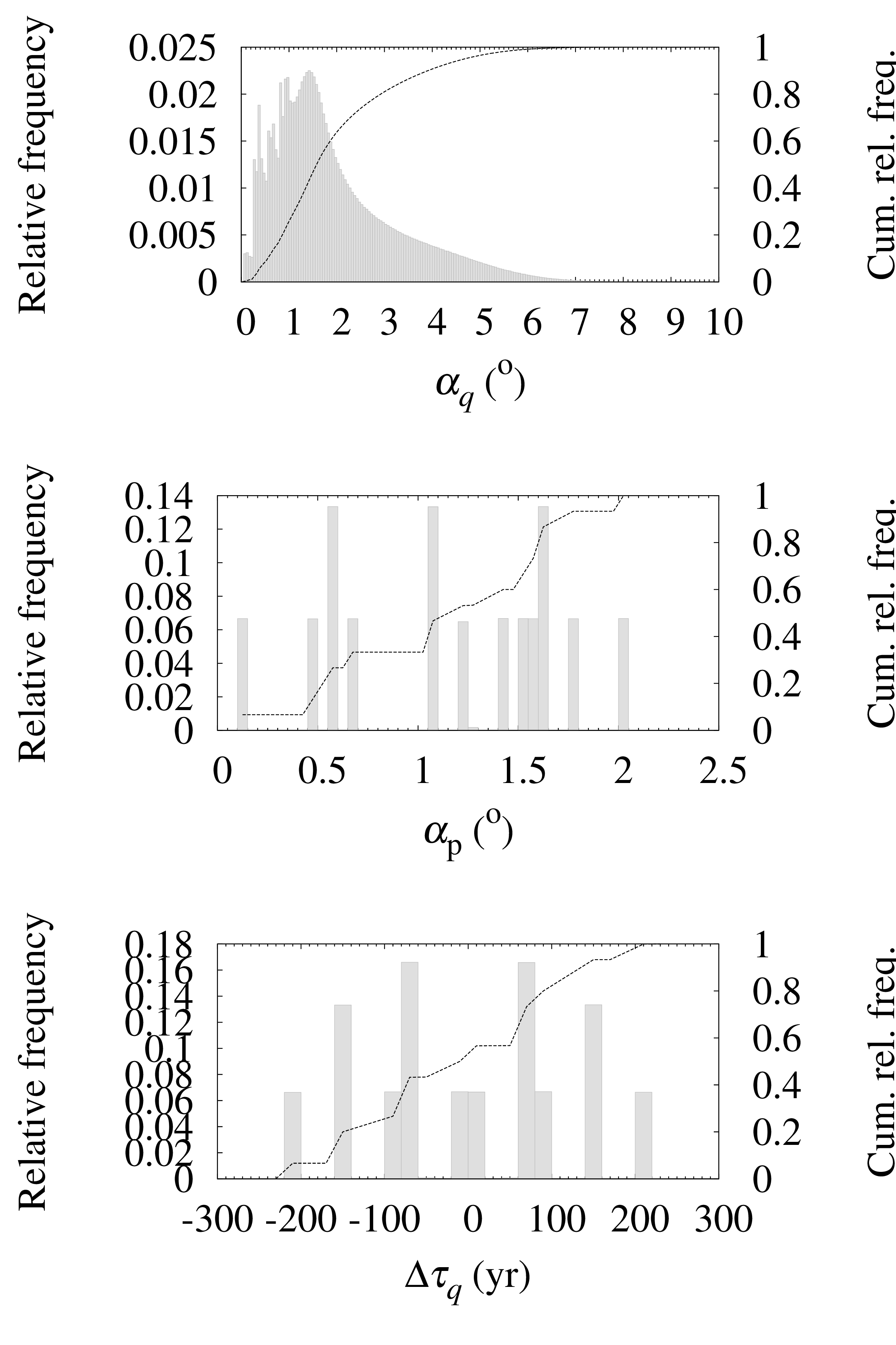}
         \includegraphics[width=0.33\linewidth]{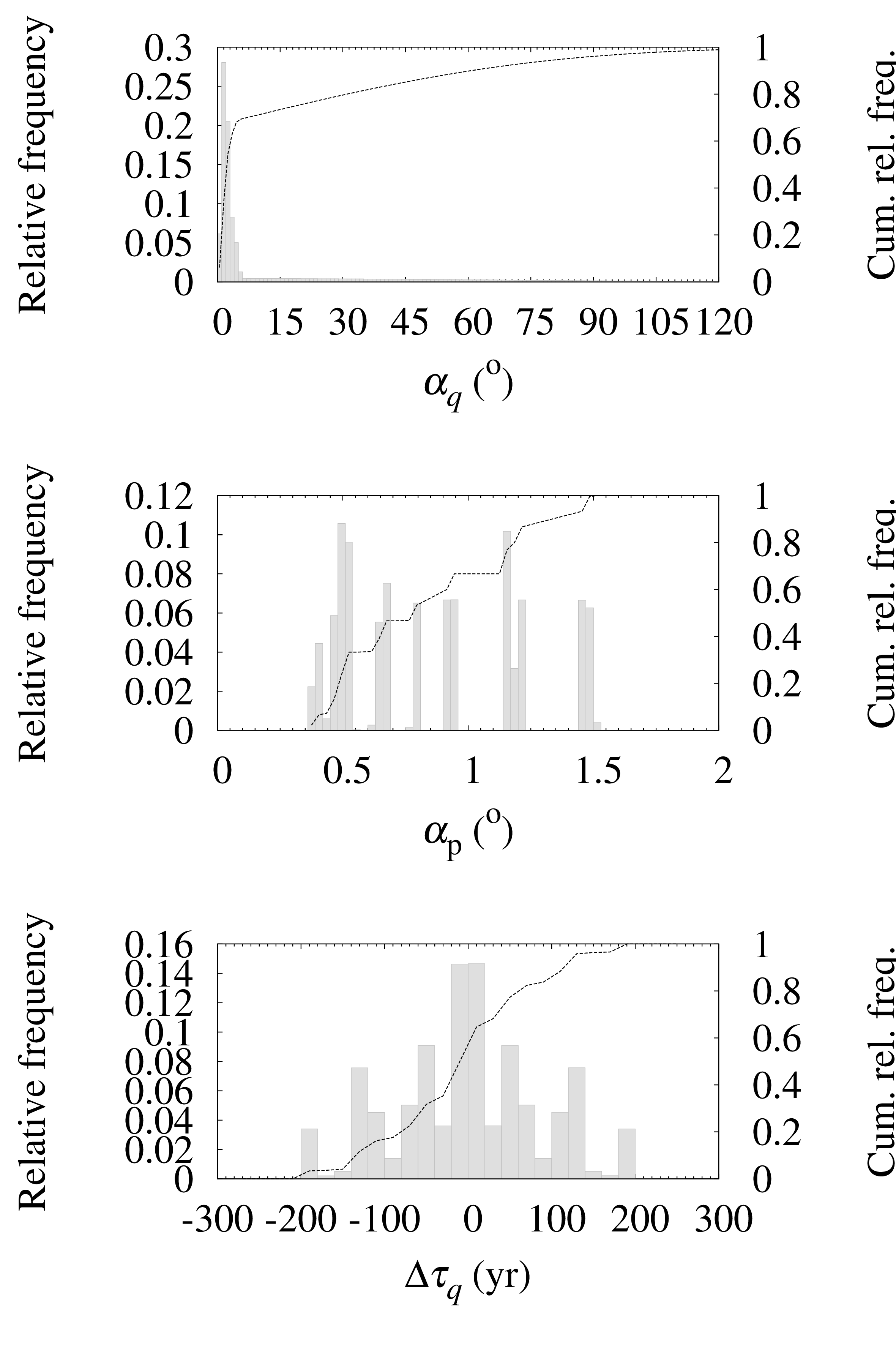}
         \caption{Distributions of possible angular separations between perihelia (top panels) and poles (middle panels), and differences in
                  time of perihelion passage (bottom panels) of the orbits of pairs of TNOs associated with (134860) 
                  2000~OJ$_{67}$--2001~UP$_{18}$ (left-hand panels), 2003~UT$_{291}$--2004~VB$_{131}$ (central panels), and 
                  2002~CU$_{154}$--2005~CE$_{81}$ (right-hand panels).
                 }
         \label{families0}
      \end{figure*}

        Our preliminary search indicates that the pair with the highest level of orbital coherence (see Table~\ref{candidates}) is the one 
        composed of (134860) 2000~OJ$_{67}$ (Buie et al. 2000) and 2001~UP$_{18}$ (Wasserman et al. 2002) with $\alpha_{\rm p}=0\fdg1825\pm0\fdg0010$, 
        $\alpha_q=0\fdg6\pm0\fdg4$ and $|\Delta\tau_{q}|=21\,361\pm145$~d. The statistical significance of the pair 134860--2001~UP$_{18}$ 
        has been estimated by computing the probability of finding a pair with lower values of both $\alpha_q$ and $\alpha_{\rm p}$ among 
        the 10$^{7}$ random pairs of virtual objects generated to single this one out; the value is $1.36\pm0.22 \times 10^{-6}$. In order 
        to evaluate the reliability of this result, we have generated a control sample of virtual TNOs with identical values of means and 
        standard deviations but assigning random (i.e. uniformly distributed) values of $\mathit{\Omega}$ and $\omega$; for this control 
        sample, the probability of finding a pair of TNOs as correlated as 134860--2001~UP$_{18}$ or better is $3.4\pm2.8 \times 10^{-7}$ 
        (mean values and standard deviations from 10 sets of experiments). Assuming that the input sample is reasonably unbiased, this 
        result may be interpreted as a confirmation that this pair might be statistically significant (at the 3.7$\sigma$ level). TNO 134860 
        is a binary itself (Stephens \& Noll 2006; Grundy et al. 2009), part of the classical trans-Neptunian belt, and one cannot avoid 
        speculating that 2001~UP$_{18}$ may be a former tertiary companion of 134860. The spectral slope or gradient of the primary of 
        134860 is 26$\pm$3\%/100~nm and the combined value for both primary and secondary is 34\%/100~nm (Hainaut, Boehnhardt \& Protopapa 
        2012); on the other hand, the spectral slope of 2001~UP$_{18}$ is 16$\pm$4\%/100~nm (Sheppard 2012). We have found that several 
        other TNOs ---e.g. (45802) 2000~PV$_{29}$, 2000~PW$_{29}$, 2001~FL$_{193}$,\footnote{This TNO has only six observations with an arc 
        length of 5\,139 d. The orbit determination provided by the Minor Planets Center is quite different from the one computed by JPL's 
        SSDG and used here.} 2001~OK$_{108}$, 2004~KF$_{19}$, 2006~HB$_{123}$, 2007~DS$_{101}$ or 2013~GX$_{136}$--- have relevant angular 
        separations within few degrees of 134860, so the presence of a collisional family is fairly likely. Fig.~\ref{families0}, left-hand 
        panels, shows the distributions in $\alpha_{\rm p}$, $\alpha_q$ and $\Delta\tau_{q}$ for the 10 candidates to constitute a 
        collisional family; the distributions of possible angular separations between perihelia and poles for possible members of this 
        dynamically coherent group certainly resemble what is seen in Fig.~\ref{disruption}, left-hand and central panels. The average 
        values and standard deviations of $a$, $e$, $i$, $\mathit{\Omega}$ and $\omega$ for the 10 candidate members are 45$\pm$2~au, 
        0.08$\pm$0.07, 1\fdg3$\pm$0\fdg5, 124\degr$\pm$83\degr and 146\degr$\pm$63\degr. On the other hand, the values of $J-H$ of 45802 
        and 2001~OK$_{108}$ are 0.43$\pm$0.29 and 0.43$\pm$0.28, respectively, which compare well with that of 134860, 0.31$\pm$0.16 
        (Hainaut et al. 2012). If this group of objects forms a collisional family, the event that might have triggered its formation should 
        have occurred less than a few Myr ago (compare Fig.~\ref{disruption} and Fig.~\ref{families0}, left-hand panels). Given the fact 
        that the values of $a$ span the range 42.7--48.5~au, the proposed TNO family includes objects from both the kernel and the stirred 
        components of the trans-Neptunian belt as characterized by Petit et al. (2011). 
%
%
      \begin{table*}
        \centering
        \fontsize{8}{11pt}\selectfont
        \tabcolsep 0.15truecm
        \caption{Orbital elements with 1$\sigma$ uncertainties of statistically significant pairs of TNOs. For each pair, the values of  
                 $\alpha_{\rm p}$, $\alpha_q$ and $|\Delta\tau_{q}|$ are listed. The orbital solutions have been computed at epoch 
                 JD 2458000.5 that corresponds to 00:00:00.000 TDB on 2017 September 4, J2000.0 ecliptic and equinox. Source: JPL's SSDG 
                 SBDB. 
                }
        \begin{tabular}{lrrrrrrrrrrrr}
          \hline
             Object                & $a$ (au) & $\sigma_a$ (au) & $e$    & $\sigma_e$ & $i$ (\degr) & $\sigma_i$ (\degr)  
                                   & $\mathit{\Omega}$ (\degr) & $\sigma_{\mathit{\Omega}}$ (\degr) & $\omega$ (\degr) & $\sigma_{\omega}$ (\degr) 
                                   & $\tau_{q}$ (JD) & $\sigma_{\tau_{q}}$ (d) \\ 
          \hline
           (134860) 2000~OJ$_{67}$ & 42.762   & 0.011           & 0.02304 & 0.00012   & 1.1147      & 0.0002
                                   & 96.795           & 0.014                     & 145.4            & 0.5 
                                   & 2430267    & 124                 \\ 
                    2001~UP$_{18}$ & 47.55    & 0.02            & 0.0819  & 0.0002    & 1.17105     & 0.00012
                                   & 105.51           & 0.05                      & 136.4            & 0.3 
                                   & 2408902    & 72                  \\
          \hline
            \multicolumn{13}{c}{$\alpha_{\rm p}=0\fdg1825\pm0\fdg0010$ \ \ $\alpha_q=0\fdg6\pm0\fdg4$ \ \ $|\Delta\tau_{q}|=21\,361\pm145$~d} \\
          \hline
                    2003~UT$_{291}$& 42.62    & 0.05            & 0.0658  & 0.0009    & 1.6126      & 0.0010
                                   & 51.152           & 0.003                     & 203.8            & 0.6 
                                   & 2410920    & 142                 \\ 
                    2004~VB$_{131}$& 43.99    & 0.02            & 0.0757  & 0.0002    & 1.7502      & 0.0006
                                   & 50.384           & 0.003                     & 205.3            & 0.4 
                                   & 2412649    & 116                 \\
          \hline
            \multicolumn{13}{c}{$\alpha_{\rm p}=0\fdg1394\pm0\fdg0012$ \ \ $\alpha_q=0\fdg9\pm0\fdg6$ \ \ $|\Delta\tau_{q}|=1726\pm184$~d}    \\
          \hline
                    2002~CU$_{154}$& 43.93    & 0.02            & 0.0628  & 0.0003    & 3.3524      & 0.0004
                                   & 108.923          & 0.005                     & 41.72            & 0.14
                                   & 2456119    & 36                  \\
                    2005~CE$_{81}$ & 42.87    & 0.02            & 0.0501  & 0.0003    & 3.0848      & 0.0008
                                   & 103.856          & 0.009                     & 45.7             & 0.3
                                   & 2457567    & 87                  \\
          \hline
            \multicolumn{13}{c}{$\alpha_{\rm p}=0\fdg3903\pm0\fdg0007$ \ \ $\alpha_q=1\fdg1\pm0\fdg4$ \ \ $|\Delta\tau_{q}|=1451\pm95$~d}     \\
          \hline
                    2000~FC$_{8}$  & 44.23    & 0.03            & 0.0584  & 0.0007    & 0.7157      & 0.0005 
                                   & 332.78           & 0.03                      & 25.4             & 0.4 
                                   & 2502959    & 178                 \\
                    2000~GX$_{146}$& 44.41    & 0.05            & 0.0073  & 0.0008    & 0.6822      & 0.0005 
                                   & 345.10           & 0.03                      & 15.0             & 2.0 
                                   & 2502905    & 605                 \\
          \hline
            \multicolumn{13}{c}{$\alpha_{\rm p}=0\fdg1536\pm0\fdg0005$ \ \ $\alpha_q=2\fdg3\pm1\fdg6$ \ \ $|\Delta\tau_{q}|=513\pm390$~d} \\
          \hline
                    2003~HF$_{57}$ & 39.68    & 0.05            & 0.1979  & 0.0007    & 1.42119     & 0.0010
                                   & 48.14            & 0.02                      & 124.6            & 0.2 
                                   & 2448299   & 32                  \\
                    2013~GG$_{137}$& 39.66    & 0.03            & 0.141   & 0.003     & 2.3905      & 0.0009      
                                   & 96.11            & 0.04                      & 75.7             & 1.2  
                                   & 2448326   & 197                 \\ 
          \hline
            \multicolumn{13}{c}{$\alpha_{\rm p}=1\fdg7844\pm0\fdg0012$ \ \ $\alpha_q=1\fdg8\pm0\fdg6$ \ \ $|\Delta\tau_{q}|=160\pm124$~d} \\
          \hline
           (135571) 2002~GG$_{32}$ & 55.88    & 0.04            & 0.3583 & 0.0005     & 14.6596     & 0.0013
                                   & 35.6950          & 0.0012                    & 230.73           & 0.12
                                   & 2460137  & 21 \\
           (160148) 2001~KV$_{76}$ & 70.60    & 0.12            & 0.5130 & 0.0007     & 15.2935     & 0.0010 
                                   & 39.5664          & 0.0003                    & 228.73           & 0.07 
                                   & 2461218  & 10 \\
          \hline
            \multicolumn{13}{c}{$\alpha_{\rm p}=1\fdg1840\pm0\fdg0009$ \ \ $\alpha_q=1\fdg75\pm0\fdg14$ \ \ $|\Delta\tau_{q}|=1081\pm23$~d} \\
          \hline
                   2005~GX$_{206}$ & 70.451   & 0.006           & 0.52982 & 0.00004 & 9.32245 & 0.00006
                                   & 126.0271 & 0.0012     & 90.065 &    0.004    
                                   & 2459658.2 & 0.5 \\
                   2015~BD$_{519}$ & 55.46    & 0.04            & 0.3482  & 0.0007  & 10.3700 & 0.0003
                                   & 120.5676 & 0.0003     & 94.424 &  0.014  
                                   & 2476148 & 12   \\
          \hline
            \multicolumn{13}{c}{$\alpha_{\rm p}=1\fdg4021\pm0\fdg0003$ \ \ $\alpha_q=1\fdg434\pm0\fdg010$ \ \ $|\Delta\tau_{q}|=16\,490\pm12$~d} \\
          \hline
        \end{tabular}
        \label{candidates}
      \end{table*}
%
%
%
%
%
%

        A similar case is found for the pair 2003~UT$_{291}$--2004~VB$_{131}$ (see Table~\ref{candidates}) which has 
        $\alpha_{\rm p}=0\fdg1394\pm0\fdg0012$, $\alpha_q=0\fdg9\pm0\fdg6$ and $|\Delta\tau_{q}|=1726\pm184$~d. Following the approach 
        discussed above, the probability of finding a pair with lower values of both $\alpha_q$ and $\alpha_{\rm p}$ is $4.6\pm1.4 \times 
        10^{-7}$, while the value from an equivalent sample with random $\mathit{\Omega}$ and $\omega$ is $9.5\pm5.9 \times 10^{-7}$. In 
        this case, it is not possible to argue that the pair is statistically significant. TNOs 2003~UT$_{291}$ (Wasserman et al. 2004) and 
        2004~VB$_{131}$ (Petit et al. 2011) may also be dynamically coherent with (33001) 1997~CU$_{29}$, 1999~OC$_{4}$, 2000~PA$_{30}$ and 
        2003~HG$_{57}$. TNO 33001 shows a very red spectrum with absorption due to water ice probably in amorphous state (Barucci et al. 
        2000); 2003~HG$_{57}$ is a known binary with a somewhat neutral spectrum (Fraser, Brown \& Glass 2015). Fig.~\ref{families0}, 
        central panels, shows the relevant distributions. The average values and standard deviations of $a$, $e$, $i$, $\mathit{\Omega}$ and 
        $\omega$ for the six objects are 44$\pm$2~au, 0.09$\pm$0.05, 1\fdg4$\pm$0\fdg6, 25\degr$\pm$38\degr, and 231\degr$\pm$37\degr. This 
        group, if real, is clearly much younger than the one linked to the pair 134860--2001~UP$_{18}$; the spreads in $\alpha_q$ and 
        $\alpha_{\rm p}$ are narrower (compare left-hand and central panels in Fig.~\ref{families0}), as are those in $\mathit{\Omega}$ 
        and $\omega$. Again, the set of objects includes members of both the kernel and the stirred components. 
        
        Another interesting pair is the one made of 2002~CU$_{154}$ (Millis et al. 2002) and 2005~CE$_{81}$ (Petit et al. 2011) which has 
        $\alpha_{\rm p}=0\fdg3903\pm0\fdg0007$, $\alpha_q=1\fdg1\pm0\fdg4$ and $|\Delta\tau_{q}|=1451\pm95$~d (see Table~\ref{candidates}). 
        The probability of finding a pair with lower values of both $\alpha_q$ and $\alpha_{\rm p}$ is $6.2\pm0.5 \times 10^{-6}$, while the 
        value from an equivalent sample with random $\mathit{\Omega}$ and $\omega$ is $8.6\pm1.4 \times 10^{-6}$. Again, it is not possible 
        to argue that the pair is statistically significant. Other TNOs that may be dynamically correlated with this pair are 2001~FK$_{193}$, 
        2003~QY$_{90}$, 2011~BV$_{163}$ and 2015~FS$_{345}$. TNO 2003~QY$_{90}$ is a known binary (Grundy et al. 2011). Fig.~\ref{families0}, 
        right-hand panels, shows the relevant distributions that resemble those of 134860--2001~UP$_{18}$. The average values and 
        standard deviations of $a$, $e$, $i$, $\mathit{\Omega}$ and $\omega$ for the six objects are 44$\pm$2~au, 0.08$\pm$0.03, 
        3\fdg6$\pm$0\fdg5, 102\degr$\pm$7\degr, and 53\degr$\pm$17\degr. This family, if real, may be as young or even younger than the one 
        linked to the pair 2003~UT$_{291}$--2004~VB$_{131}$. As in the previous two candidate families, the presumed members belong to the 
        kernel and the stirred components of the dynamically cold trans-Neptunian belt. 
%
%
      \begin{figure*}
        \centering
         \includegraphics[width=0.49\linewidth]{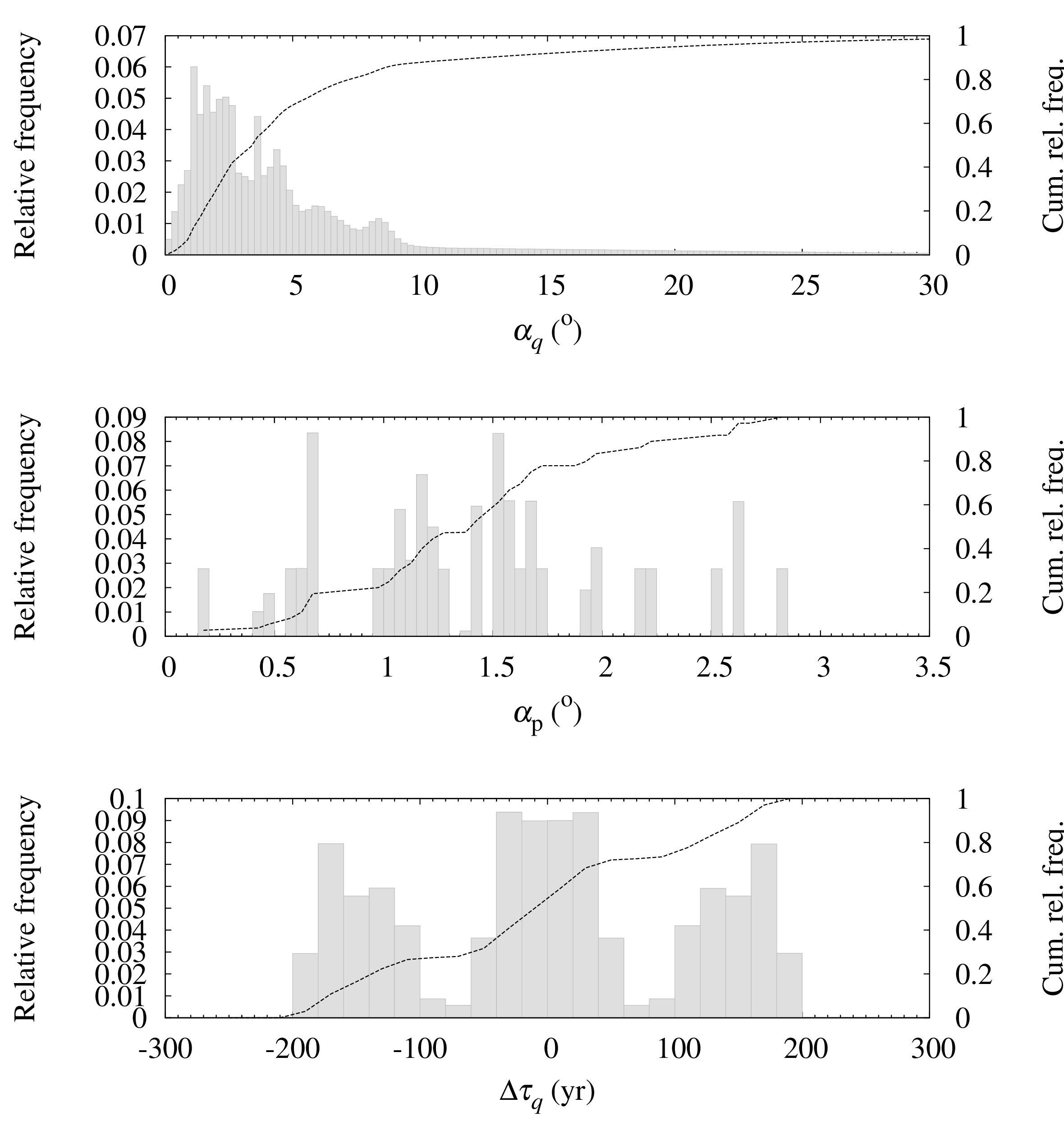}
         \includegraphics[width=0.49\linewidth]{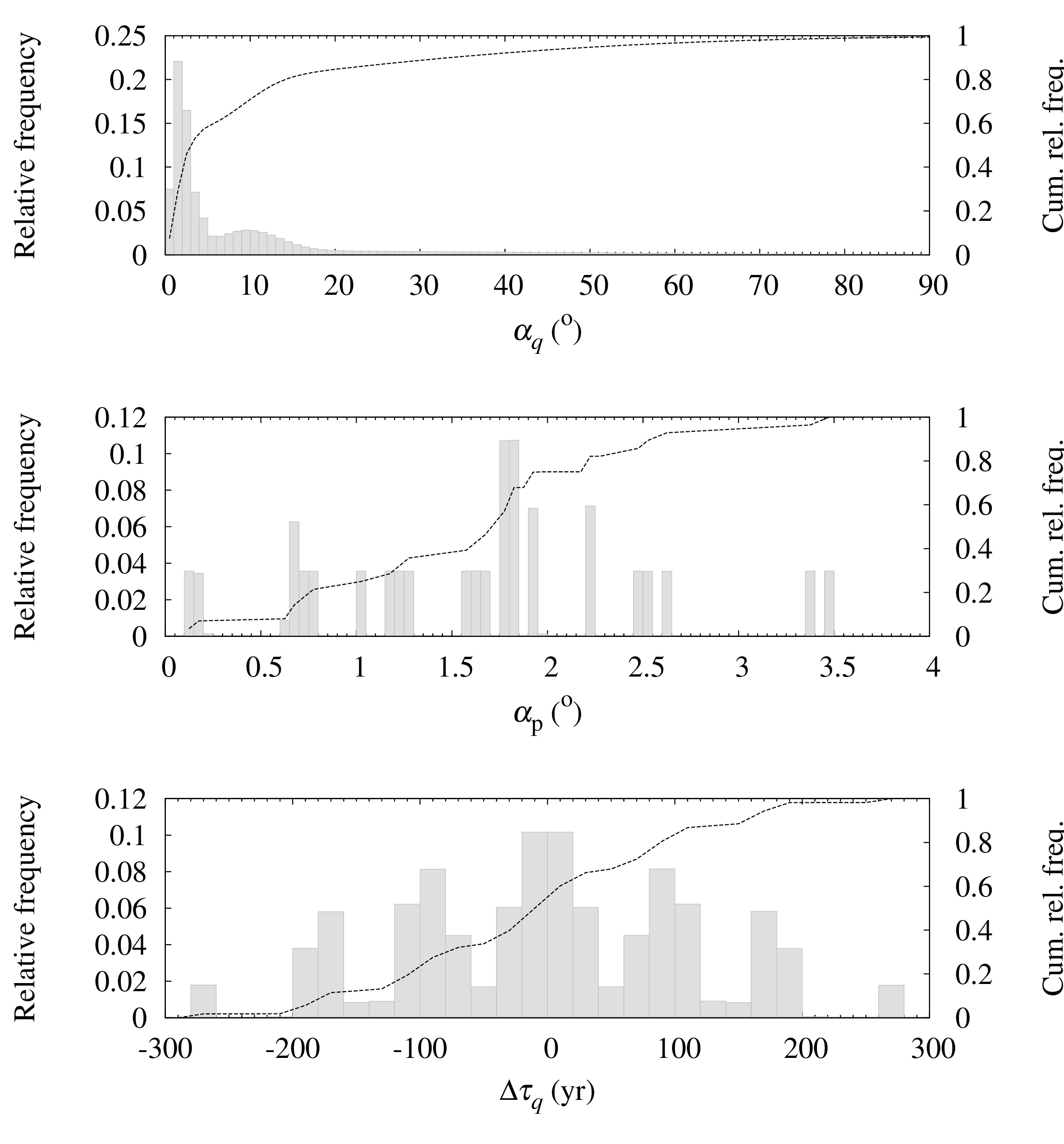}
         \caption{Distributions of possible angular separations between perihelia (top panels) and poles (middle panels), and differences in
                  time of perihelion passage (bottom panels) of the orbits of pairs of TNOs associated with 2000~FC$_{8}$--2000~GX$_{146}$ 
                  (left-hand panels) and 2003~HF$_{57}$--2013~GG$_{137}$ (right-hand panels).
                 }
         \label{families}
      \end{figure*}
%
%
        
        We also recover the pair 2000~FC$_{8}$--2000~GX$_{146}$ which has $\alpha_{\rm p}=0\fdg1536\pm0\fdg0005$, $\alpha_q=2\fdg3\pm1\fdg6$ 
        and $|\Delta\tau_{q}|=513\pm390$~d (see Table~\ref{candidates}). These two TNOs are part of the candidate collisional family 
        originally proposed by Chiang (2002), but apparently later retracted by Chiang et al. (2003), although the original result has been 
        somewhat vindicated by Petit et al. (2011) and Marcus et al. (2011). The probability of finding a pair with lower values of both 
        $\alpha_q$ and $\alpha_{\rm p}$ is $3.1\pm0.4 \times 10^{-6}$, while the value from an equivalent sample with random 
        $\mathit{\Omega}$ and $\omega$ is $1.8\pm0.7 \times 10^{-6}$. Therefore, this pair might be statistically significant (at the 
        1.9$\sigma$ level). Other TNOs that belong to the kernel or the stirred components and that may be dynamically related to this pair 
        are (15760) 1992~QB$_{1}$, 2000~QN$_{251}$, 2002~CY$_{154}$, 2002~PV$_{170}$, 2005~VA$_{123}$, 2010~TG$_{192}$ and 2015~GS$_{50}$. 
        Using these TNOs we obtain the distributions in Fig.~\ref{families}, left-hand panels, that clearly resemble what is expected 
        of an event of progressive fragmentation like that of 73P (see Fig.~\ref{disruption}, left-hand panels). The average values and 
        standard deviations of $a$, $e$, $i$, $\mathit{\Omega}$ and $\omega$ for the nine objects are 44$\pm$2~au, 0.10$\pm$0.09, 
        1\fdg2$\pm$0\fdg6, 221\degr$\pm$143\degr, and 139\degr$\pm$141\degr. Not counting Charon, 15760 was the second TNO to be discovered 
        after the dwarf planet Pluto (Jewitt \& Luu 1993; Luu, Jewitt \& Marsden 1993). This candidate family is certainly older than those 
        associated with the pairs 2003~UT$_{291}$--2004~VB$_{131}$ or 2002~CU$_{154}$--2005~CE$_{81}$ and the presence of secondary maxima 
        in Fig.~\ref{families}, left-hand, top panel, suggests that multiple disruption events may have contributed to create this group.  

        Yet another pair of interest that seems to be signalling the presence of an additional collisional family is composed of 2003~HF$_{57}$ 
        (Kavelaars et al. 2009) and 2013~GG$_{137}$ (Bannister et al. 2016) ---see Table~\ref{candidates}--- with $\alpha_{\rm p}=1\fdg7844\pm0\fdg0012$, 
        $\alpha_q=1\fdg8\pm0\fdg6$ and $|\Delta\tau_{q}|=160\pm124$~d. The probability of finding a pair with lower values of both $\alpha_q$ 
        and $\alpha_{\rm p}$ is $2.76\pm0.02 \times 10^{-4}$, while the value from an equivalent sample with random $\mathit{\Omega}$ and 
        $\omega$ is $2.19\pm0.10 \times 10^{-4}$. Therefore, this pair might be statistically significant (at the 5.6$\sigma$ level). Other 
        TNOs that could be dynamically related to this pair are 1998~WY$_{31}$, 2000~ON$_{67}$, 2001~FL$_{193}$, 2003~QX$_{90}$, 
        2005~JP$_{179}$ and 2006~HW$_{122}$. The presence of one object, 2001~FL$_{193}$, in common with the group associated with the pair 
        134860--2001~UP$_{18}$ may hint at a possible relationship between the two groups of TNOs or, more likely, that this object is an 
        interloper ---or even that its orbit determination cannot be trusted, see footnote (3). This set of TNOs defines the distributions 
        in Fig.~\ref{families}, right-hand panels, and again the presence of secondary maxima (top panel) hints at multiple disruption 
        events as in the case of the group linked to 2000~FC$_{8}$--2000~GX$_{146}$. The average values and standard deviations of $a$, $e$, 
        $i$, $\mathit{\Omega}$ and $\omega$ for the eight objects are 43$\pm$2~au, 0.09$\pm$0.06, 2\fdg1$\pm$0\fdg8, 73\degr$\pm$31\degr, 
        and 100\degr$\pm$31\degr. These values suggest that this group could be as young as those associated with the pairs 
        2003~UT$_{291}$--2004~VB$_{131}$ or 2002~CU$_{154}$--2005~CE$_{81}$, but the distributions in Fig.~\ref{families}, right-hand 
        panels, indicate that it could be older than the collisional family linked to the pair 2000~FC$_{8}$--2000~GX$_{146}$.
        
        Several other relevant pairs are slightly less significant and may be linked to additional families, but they will not be discussed 
        here. The statistical significance of individual pairs is expected to be strongly contingent on the quality and completeness of the 
        input data; if there is a finite number of collisional families, the probabilities may depend on the degree of completeness in which
        each family is represented within the available data. Petit et al. (2011) pointed out that the presence of collisional families like 
        the one probably linked to the pair 2000~FC$_{8}$--2000~GX$_{146}$ and previously discussed by Chiang (2002) is supported by the 
        observational data, and that grazing impacts like those discussed by Leinhardt et al. (2010) can produce low-speed families. If the 
        dynamically coherent groups discussed here are confirmed to be genetic families, then it may be possible that such grazing 
        collisions could be actively grinding relatively large TNOs (of hundreds of km) up. An alternative scenario would place the source 
        of the observed dynamical coherence in mean motion and secular resonances, but it is unclear how this might work. However, the data 
        suggest that the stirred cold population might just be a splitted counterpart of the kernel component as objects in the 2:1, 3:2 
        mean motion resonances with Neptune and possibly even a scattered one ---see footnote (3)--- appear to be correlated to cold disc 
        kernel objects.

     \subsection{Semimajor axes in the range 50--150~au}
        In general, the quality of the orbits of TNOs with $a\in(50,~150)$~au is lower than that of the objects analysed in the previous 
        section and the available samples are more likely to be affected by observational biases and incompleteness issues. Most of the TNOs 
        in this region have dynamically hot orbits, with a very small fraction of dynamically cold objects. Volk \& Malhotra (2017) have 
        presented robust statistical evidence that the mean plane of the trans-Neptunian belt is warped in a manner that an inclined, 
        low-mass (probably Mars-sized), unseen planet at an average distance from the Sun of around 60~au could be responsible for the 
        warping. Such a perturber may also induce (or help wipe out) orbital coherence.

        One pair of correlated objects is (135571) 2002~GG$_{32}$, a red object with a spectral slope of 34$\pm$3\%/100~nm (Sheppard 2012), 
        and (160148) 2001~KV$_{76}$ (Elliot et al. 2005) which has $\alpha_{\rm p}=1\fdg1840\pm0\fdg0009$, $\alpha_q=1\fdg75\pm0\fdg14$ and 
        $|\Delta\tau_{q}|=1081\pm23$~d. Using the approach discussed in the previous section, the probability of finding a pair with lower 
        values of both $\alpha_q$ and $\alpha_{\rm p}$ is $7.4\pm1.1 \times 10^{-6}$, while the value from an equivalent sample with random 
        $\mathit{\Omega}$ and $\omega$ is $2.1\pm1.5 \times 10^{-5}$. Such a large difference in the values of the probabilities must be the 
        result of strong non-uniformity in the distributions of the observed values of $\mathit{\Omega}$ and $\omega$ that could be due to 
        observational bias or selection effects, or (perhaps more likely) induced by an unseen perturber. The two orbits exhibit a high 
        degree of coherence in terms of the angular elements and $\tau_q$, but their values of $a$ and $e$ are very different (see 
        Table~\ref{candidates}). This is consistent with binary dissociation induced by a close encounter with a massive body, not 
        fragmentation (see e.g. de la Fuente Marcos et al. 2017). A similar case is found for the pair 2005~GX$_{206}$ (Gibson et 
        al. 2016a) and 2015~BD$_{519}$ (Gibson et al. 2016c) with $\alpha_{\rm p}=1\fdg4021\pm0\fdg0003$, $\alpha_q=1\fdg434\pm0\fdg010$ and 
        $|\Delta\tau_{q}|=16\,490\pm12$~d. The probability of finding a pair with lower values of both $\alpha_q$ and $\alpha_{\rm p}$ is 
        $7.6\pm0.9 \times 10^{-6}$, while the value from an equivalent sample with random $\mathit{\Omega}$ and $\omega$ is $1.8\pm1.4 
        \times 10^{-5}$. As in the previous case, the very different values of the probabilities might signal the presence of a present-day,
        relatively massive perturber. There are other pairs with similar statistical significance. 

        Another pair of potentially interesting objects (not shown in Table~\ref{candidates}) is the one composed of 2012~OL$_{6}$ (Sheppard 
        \& Trujillo 2016) and 2014~WS$_{510}$ (Gibson et al. 2016b) which has $\alpha_{\rm p}=0\fdg185\pm0\fdg008$, $\alpha_q=8\degr\pm6\degr$ 
        and $|\Delta\tau_{q}|=49\,002\pm3\,884$~d. Other TNOs that might be dynamically related to this pair are 2002~CZ$_{248}$ and 
        2013~AR$_{183}$. TNOs 2012~OL$_{6}$, 2014~WS$_{510}$ and 2002~CZ$_{248}$ have average values and standard deviations of $a$, $e$, 
        $i$, $\mathit{\Omega}$ and $\omega$ of 54.1$\pm$1.1~au, 0.35$\pm$0.05, 8\degr$\pm$2\degr, 153\degr$\pm$15\degr, and 286\degr$\pm$22\degr;
        in contrast, 2013~AR$_{183}$ has $a$=71.7~au. As a reference, the candidate pair in Rabinowitz et al. (2011), (471151) 
        2010~FD$_{49}$--(471152) 2010~FE$_{49}$, has $\alpha_{\rm p}=1\fdg89243\pm0\fdg00013$, $\alpha_q=18\fdg735\pm0\fdg009$ and 
        $|\Delta\tau_{q}|=4019.8\pm1.4$~d.

     \subsection{Beyond 150~au}
        If we focus on extreme Centaurs and trans-Neptunian objects ($a$$>$150~au), the pair with the smallest angular separations is 
        composed of 2013~FT$_{28}$ (Sheppard \& Trujillo 2016) and 2015~KG$_{163}$ (Shankman et al. 2017) with $\alpha_{\rm p}=3\fdg379\pm0\fdg004$, 
        $\alpha_q=7\fdg4\pm0\fdg2$ and $|\Delta\tau_{q}|=13\,640\pm22$~d, although their $a$ and $e$ are quite different. Its probability 
        from Fig.~\ref{distributions}, right-hand panels, is $<$0.0003 ---that of having smaller values of $\alpha_{\rm p}$ and 
        $\alpha_q$. This pair could be similar to (474640) 2004~VN$_{112}$--2013~RF$_{98}$ studied by de Le{\'o}n et al. (2017) although in 
        this case the values of $a$ and $e$ are similar. The TNO pair 474640--2013~RF$_{98}$ has $\alpha_{\rm p}=4\fdg055\pm0\fdg003$,
        $\alpha_q=14\fdg2\pm0\fdg6$ and $|\Delta\tau_{q}|=101\pm74$~d. Another pair of extreme TNOs with similar orbital elements is the
        one composed of 2002~GB$_{32}$ (Meech et al. 2004) and 2003~HB$_{57}$ (Kavelaars et al. 2009) which has $\alpha_{\rm p}=5\fdg4543\pm0\fdg0006$, 
        $\alpha_q=7\fdg39\pm0\fdg05$ and $|\Delta\tau_{q}|=1856\pm11$~d.

  \section{Conclusions}
     In this paper, we have searched for dynamically correlated minor bodies in the outer Solar system. A novel technique that uses the 
     angular separations of orbital poles and perihelia, together with the differences in time of perihelion passage has been described and
     applied to find statistically significant pairs and groups. In summary: 
     \begin{enumerate}[(i)]
        \item We provide further evidence that confirms the reality of the candidate collisional family of TNOs associated with the pair 
              2000~FC$_{8}$--2000~GX$_{146}$ and originally proposed by Chiang (2002).
        \item We find four new possible collisional families of TNOs associated with the pairs (134860) 2000~OJ$_{67}$--2001~UP$_{18}$ 
              2003~UT$_{291}$--2004~VB$_{131}$, 2002~CU$_{154}$--2005~CE$_{81}$ and 2003~HF$_{57}$--2013~GG$_{137}$.
        \item We find a number of unbound TNOs that may have a common origin, the most significant ones are: 
              (135571) 2002~GG$_{32}$--(160148) 2001~KV$_{76}$ and 2005~GX$_{206}$--2015~BD$_{519}$.
     \end{enumerate}
     Our results suggest that disruptions and dissociations of minor bodies at tens or even hundreds of astronomical units from the Sun 
     could be as common as those taking place much closer to us. Future spectroscopic observations may help in confirming the dynamical 
     correlations found here.

  \section*{Acknowledgements}
     We thank the anonymous referee for helpful and prompt reports that included a number of critical and constructive enquiries, A.~I. 
     G\'omez de Castro, I. Lizasoain and L. Hern\'andez Y\'a\~nez of the Universidad Complutense de Madrid (UCM) for providing access to 
     computing facilities. This work was partially supported by the Spanish `Ministerio de Econom\'{\i}a y Competitividad' (MINECO) under 
     grant ESP2014-54243-R. Part of the calculations and the data analysis were completed on the EOLO cluster of the UCM, and we thank S. 
     Cano Als\'ua for his help during this stage. EOLO, the HPC of Climate Change of the International Campus of Excellence of Moncloa, is 
     funded by the MECD and MICINN. This is a contribution to the CEI Moncloa. In preparation of this paper, we made use of the NASA 
     Astrophysics Data System, the ASTRO-PH e-print server, and the MPC data server.

  \bsp
  \label{lastpage}
\end{document}